\documentclass{webofc}
\usepackage[varg]{txfonts}   % Web of Conferences font
\begin{document}
\title{Intergalactic electromagnetic cascades \\ in the magnetized Universe \\ as a tool of astroparticle physics}

\author{\firstname{Timur} \lastname{Dzhatdoev}\inst{1,2}\fnsep\thanks{\email{timur1606@gmail.com}} \and
        \firstname{Emil} \lastname{Khalikov}\inst{1} \and
        \firstname{Anna} \lastname{Kircheva}\inst{1,3} \and
        \firstname{Egor} \lastname{Podlesnyi}\inst{1,3} \and \\
        \firstname{Anastasia} \lastname{Telegina} \inst{3}}

\institute{Skobeltsyn Institute of Nuclear Physics, Moscow State University, Moscow, 119991 Russia 
\and
Institute for Cosmic Ray Research, University of Tokyo, 5-1-5 Kashiwanoha, Kashiwa, Japan
\and
Faculty of Physics, Moscow State University, Moscow, 119991 Russia}

\abstract{We review the physics of intergalactic electromagnetic cascades in the presence of the extragalactic magnetic field (EGMF). Various regimes of intergalactic electromagnetic cascades are considered depending on the number of cascade generations, the value of the cascade electron deflection angle, and the relations between the EGMF coherence length, typical cascade $\gamma$-ray mean free path, and electron energy loss length. We also review contemporary constraints on the EGMF parameters and explore the sensitivity of various $\gamma$-ray instruments to the EGMF parameters.}
\maketitle
\section{Introduction} \label{sec:intro}

Primary $\gamma$-rays from extragalactic sources may be absorbed on extragalactic background light (EBL) \cite{nikishov_1962,gould_schreder_1967} and cosmic microwave background (CMB) photons \cite{jelley_1966} through the $\gamma\gamma \rightarrow e^{+}e^{-}$ pair production (PP) process. For sufficiently distant sources with redshift $z>$0.1, the optical depth $\tau$ of the PP process exceeds unity for the primary energy $E_{0}>$1 TeV. Therefore, above some energy $E(\tau=1)$ (usually called ``the gamma-ray horizon'' and defined by the so-called Fazio-Stecker relation \cite{fazio_stecker_1970}) primary spectra of extragalactic $\gamma$-ray sources are strongly distorted. This effect, evidently, creates appreciable obstacles for the direct study of distant very high energy (VHE, $E>$100 GeV) $\gamma$-ray emitters.

On the other hand, $\gamma$-ray absorption could be an asset (e.g. \cite{neronov_semikoz_2007}, \cite{neronov_semikoz_2009}). Indeed, secondary electrons and positrons (in what follows they are called simply ``electrons'', unless it is necessary to distinguish $e^{+}$ and $e^{-}$) produce secondary (cascade) $\gamma$-rays through inverse Compton (IC) scattering. These cascade $\gamma$-rays are observable; their spectral shape may help constrain the shape of the primary $\gamma$-ray spectrum in the optically thick ($\tau>$1) energy range. Moreover, observable spectral, angular, and temporal distributions of cascade $\gamma$-rays are sensitive to parameters of the extragalactic magnetic field (EGMF). Thus, these parameters, such as the EGMF characteristic strength $B$ and coherence length $\lambda$, could be constrained using observations of extragalactic $\gamma$-ray sources.

This case exemplifies how intergalactic electromagnetic (EM) cascades may be a valuable tool of astroparticle physics. Deep understanding of the underlying physics is important in order to use this tool effectively. EM cascades in the magnetized Universe may reveal a number of qualitatively different regimes depending on the following basic parameters: the primary energy, the distance from the source to the observer, the strength and coherence length of the EGMF, etc. The most general constraints on the EGMF parameters in voids of the large scale structure (LSS) that define the ``EGMF parameter window'' are provided in section~\ref{sec:window}. In section~\ref{sec:regimes} we discuss various regimes of intergalactic EM cascades. In section~\ref{sec:egmf} we briefly review some contemporary constraints on the EGMF parameters $(B,\lambda)$. Section~\ref{sec:sensitivity} is devoted to a study of sensitivity of various $\gamma$-ray instruments to the EGMF parameters. Some other case studies performed by us were described in \cite{dzhatdoev_et_al_2017}. Finally, we provide conclusions in section~\ref{sec:conclusions}. In this work we assume $z$=0.186 and the EBL model of \cite{gilmore_et_al_2012} unless stated otherwise. This EBL model is consistent with contemporary constraints (e.g. \cite{korochkin_rubtsov_2018}). We made use of the ROOT analysis framework \cite{brun_rademakers_1997} and the MINUIT routine \cite{james_roos_1975}.

\section{The EGMF parameter window} \label{sec:window}

\begin{figure}[h]
\centering
\includegraphics[width=8cm]{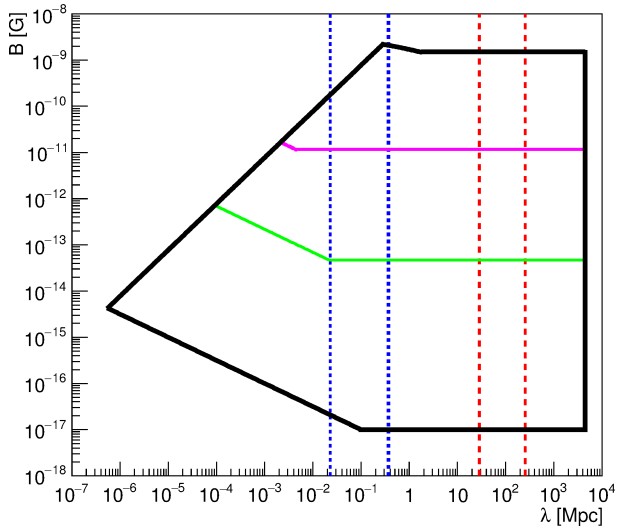}
\caption{General constraints on the EGMF parameters in LSS voids together with a range of $\gamma$-ray mean free paths (red dashed lines), a range of electron energy loss lengths (blue dashed lines), and the borders of the PH and MBC regimes (magenta and green lines, respectively). \label{fig:Regimes}}
\end{figure}

The strength and structure of the EGMF in voids is currently poorly constrained. Below we assume a simplified EGMF model that is fully described by two parameters $(B,\lambda)$. Fig. \ref{fig:Regimes} shows some basic constraints on these parameters at $z$=0 following \cite{durrer_neronov_2013} (solid thick black frame). Vertical line that corresponds to the right border of the frame denotes the Hubble radius $R_{H}$. As a rule, EM cascades develop on spatial scales significantly smaller than $R_{H}$. Therefore, observables of intergalactic EM cascade are virtually independent of $\lambda$ if $\lambda>R_{H}$. Thus, while there is no firm upper limit (either observational or theoretical) on the EGMF coherence length $\lambda$ \cite{campanelli_2016}, in what follows  we assume $\lambda<R_{H}$ for convenience. On the other hand, a lower theoretical bound on $\lambda$ could be set (see \cite{durrer_neronov_2013}); it is denoted as upper-left part of the frame in fig. \ref{fig:Regimes}. Constraints from Faraday rotation (the upper part of the frame) according to \cite{pshirkov_tinyakov_urban_2016} are also shown. Finally, the lower part of the frame corresponds to the constraints from apparent non-observation of the cascade component of blazar emission \cite{vovk_et_al_2012}.

\section{Regimes of intergalactic EM cascades} \label{sec:regimes}

In the present work we mainly consider the case of $z$= 0.1--0.3 that is typical for blazars with hard primary spectra (the so-caled ``extreme TeV blazars'') that additionally have measured intensity in energy bins where $\tau>$2 (see Table 1 in \cite{dzhatdoev_et_al_2017} and \cite{bonnoli_et_al_2015}). We neglect synchrotron energy losses for electrons (which may be important for extremely high energy sources in LSS filaments, e.g. \cite{uryson_2018}) and collective (plasma) effects \cite{broderick_chang_pfrommer_2012}. We also do not consider processes beyond the Standard Model (cf. \cite{horns_meyer_2012, rubtsov_troitsky_2014}; see, however, \cite{biteau_williams_2015, dominguez_ajello_2015} and \cite{dzhatdoev_et_al_2017}). A detailed discussion of intergalacic EM cascades from primary protons \cite{uryson_1998, essey_kalashev_kusenko_beacom_2011, murase_dermer_takami_migliori_2012} is also available in \cite{dzhatdoev_et_al_2017}.

The mean free path of $\gamma$-rays ($L_{\gamma}$) and electrons ($L_{e}$) with primary energy $E_{0}<10$ EeV is smaller than 10 Mpc; therefore, EM cascades develop very fast until the threshold of the pair-production process on the CMB is reached (e.g. \cite{berezinsky_et_al_2011}). For the case of $E_{0}>$10 EeV, the mean free path of primary $\gamma$-rays on the CMB may exceed 10 Mpc \cite{bonometto_1971, bonometto_lucchin_1971}, but the total interaction rate of $\gamma$-rays also depends on the spectral shape and intensity of the universal radio background (e.g. \cite{protheroe_biermann_1996}), which are poorly known compared to the ones for the CMB and even for the EBL. We leave the case of $E_{0}>$10 EeV for future studies.

\subsection{Classification by the number of generations}

Intergalactic EM cascades may have one or more generations. In \cite{dzhatdoev_et_al_2017} we show that the set of parameters ($z=$0.186; $E_{0}<$10 TeV) corresponds to the case of the one-generation regime, when the peak energy in the spectral energy distribution (SED) of cascade $\gamma$-rays strongly ($\propto E_{0}^{2}$) depends on the primary energy. On the other hand, for the same $z$ and $E_{0}=$100 TeV--1 EeV the observable spectrum of $\gamma$-rays depends on $E_{0}$ only weakly, corresponding to the universal regime \cite{berezinsky_kalashev_2016} of many cascade generations. This energy border between the regimes shifts to higher energies with the fall of $z$. A more detailed discussion of the one-generation and universal regimes is avaiable in \cite{dzhatdoev_et_al_2017}.

\subsection{Classification by the deflection angle value}

Cascade electrons get deflected in non-zero EGMF. Depending on the typical deflection angle value $\delta$, two basic regimes of intergalactic EM cascades could be identified \cite{abramowski_et_al_2014}, namely: magnetically broadened cascade (MBC) for the case of $\delta<<$1 rad and pair halo (PH) \cite{aharonian_coppi_voelk_1994} for the case of $\delta>$1 rad. In both cases the angular distribution of observable $\gamma$-rays is broadened with respect to the primary $\gamma$-ray angular distribution; therefore, the effect of PH or MBC on the observable angular distribution may be collectively described as ``extended cascade emission''. Green solid line in fig. 1 shows the upper border (in terms of $B$) of the MBC regime for the case of primary $\gamma$-rays with energy 20 TeV assuming $\delta$= 0.1 rad. For a more detailed discussion of the geometry specific to the MBC regime see \cite{dzhatdoev_khalikov_kircheva_podlesnyi_2017}. Magenta solid line shows the lower border (again in terms of $B$) of the PH regime for primary $\gamma$-rays with energy 100 TeV, assuming $\delta$= 1 rad.

\subsection{Classification by the axial symmetry/asymmetry of the cascade}

In the MBC regime, the average length of the cascade is appoximately equal to the mean free path of $\gamma$-rays that produce the observable $\gamma$-rays (this quantity may be denoted as $L_{\gamma-(N-1)}$, where $N$ is the typical number of the last observable generation of $\gamma$-rays). The energy range of such ``last parent generation'' $\gamma$-rays is defined by two conditions: 1) they should interact with EBL photons strongly enough in order for the next generation to appear (i.e. for them typically $\tau$>1) 2) however, their secondary population of $\gamma$-rays should have optical depth below unity, otherwise these secondary $\gamma$-rays would produce another generation of the cascade. The mean free path of $\gamma$-rays with primary energy (source restframe) between 100 GeV and 300 TeV may be approximated by the following equation with typical precision of $\sim$10 \%:
\begin{equation}
L_{\gamma}(E,z)= C\cdot L_{t} \frac{E_{t}}{(1+z)^{\alpha}E}[1+k\cdot sin(a\cdot lg[(1+z)^{\beta}E]-b)] \label{eqn05},
\end{equation}
where $E_{t},L_{t}$ are fixed at (10 $TeV$, 80 $Mpc$), correspondingly, and the parameter values are $C = 0.979$, $\alpha = 2.67$, $\beta= 0.857$, $k= 0.553$, $a = 3.04$, $b= 1.28$ \cite{dzhatdoev_khalikov_kircheva_podlesnyi_2017}. Vertical red dashed lines in fig.~\ref{fig:Regimes} show the value of $L_{\gamma}$ for two energies of $\gamma$-rays: 20 TeV (left line) and 1 TeV (right line). These energies roughly correspond to the energy range of the ``last parent generation'' for the considered range of conditions. Therefore, for the case of $\lambda\gg L_{\gamma-(N-1)}$ most of intergalactic EM cascades would develop inside a practically coherent EGMF; cascade electrons and positrons would be deflected to opposite directions, thus the cascade would be axially asymmetric \cite{neronov_et_al_2010, broderick_et_al_2016}. In the opposite case of $\lambda\ll L_{\gamma-(N-1)}$ the cascade would be practically axially-symmetric. This classification, however, applies only to the MBC regime, as for the case of the PH regime the axial symmetry is partially restored due to strong deflections of cascade electrons.

\subsection{Classification by the electron propagation mode}

The characteristic electron energy loss length may be approximated with very good accuracy ($\sim$1 \%) as shown in \cite{khangulyan_aharonian_kelner_2014}. Vertical blue dashed lines in fig.~\ref{fig:Regimes} show the energy loss length of electrons $L_{E-e}$ calculated for two energies of electrons: 500 GeV (right line) and 10 TeV (left line) according to \cite{khangulyan_aharonian_kelner_2014}. In the case of $\lambda \gg L_{E-e}$ the electron loses most of its energy inside a practically homogeneous patch of the EGMF, thus experiencing regular deflection, and the deflection angle $\delta$ is roughly proportional to $L_{E-e}$. On the other hand, in the case of $\lambda \ll L_{E-e}$, the electron traverses many EGMF cells with various field directions, thus $\delta\propto \sqrt{L_{E-e}}$ \cite{neronov_semikoz_2009}.

\section{Contemporary constraints on the EGMF parameters in voids} \label{sec:egmf}

Below we briefly summarise some contemporary constraints on the EGMF parameters in voids at $z$=0. \\
1. Faraday rotation measurements at present provide only upper limits on $B$, typically aroung 1 nG \cite{blasi_et_al_1999,pshirkov_tinyakov_urban_2016}. However, the Faraday tomography approach could be useful to constrain or even measure the EGMF in the LSS, in particular, in filaments \cite{akahori_ryu_2010}. \\
2. Galaxy cluster simulations may be used to derive the strength of the EGMF in voids sufficient to produce observable magnetic fields. $B\sim$2 pG is usually deemed sufficient \cite{dolag_et_al_2005}. This could be viewed as a model-dependent upper limit on $B$ in voids. We note that some models exist that allow for $B<10^{-16}-10^{-18}$ G in voids while they are still able to produce present-day magnetic fields in galaxy clusters (e.g., \cite{hackstein_et_al_2018}). \\
3. Constraints obtained from CMB studies usually provide upper limits on $B$ of the order of 1 nG \cite{ade_et_al_2016}. However, recently it was claimed that model-dependent constraints of the order of 10-50 pG are possible to obtain \cite{jedamzik_saveliev_2018}. \\
4. Weak (<0.1 pG) magnetic fields in voids could potentially be measured using EM cascades. Up to date, there are several related techniques available: \\
a) Direct search for the MBC pattern in angular distributions of blazars \cite{abramowski_et_al_2014,chen_et_al_2015,archambault_et_al_2017}. Assuming $\lambda$= 1 Mpc, such searches are most sensitive for $B$= 1--10 fG for the case of imaging atmospheric Cherenkov telescopes (IACT) such as H.E.S.S. \cite{abramowski_et_al_2014} and VERITAS \cite{archambault_et_al_2017}, and for $B$= 0.01--1 fG for the case of the Fermi LAT telescope \cite{chen_et_al_2015}. An interesting special case is that of the asymmetric MBC \cite{neronov_et_al_2010}, \cite{broderick_et_al_2016} (i.e. large values of $\lambda>10^{2}-10^{3}$ Mpc) when additional information becomes available, providing stronger constraints on the EGMF parameters \cite{tiede_et_al_2017}. The extended emission search technique is also sensitive to the PH pattern, but in this case only lower constraints on $B$ could be derived \cite{aharonian_coppi_voelk_1994}.\\
b) A simplified method that relies solely on the spectral information inside the point spread function (PSF) of the instrument was, so far, the most popular technique \cite{neronov_vovk_2010, tavecchio_et_al_2010, taylor_vovk_neronov_2011, vovk_et_al_2012, takahashi_et_al_2012, takahashi_et_al_2013, finke_et_al_2015, arXiv:1804.08035v1}. However, in this case a part of information is lost, and the results are less robust than for the case of the MBC search method; therefore, the systematic uncertainty of this approach is large, sometimes not even allowing to exclude the hypothesis of $B$=0 \cite{arlen_et_al_2014}. Additionally, for the case of flaring sources the effect of time delay of cascade photons \cite{plaga_1995} leads to additional ambiguity \cite{dermer_et_al_2011} (see also \cite{ichiki_inoue_takahashi_2008, murase_et_al_2008}). \\
c) For the case of the helical EGMF some constraints may be obtained from diffuse $\gamma$-rays \cite{tashiro_et_al_2014}.
 
\section{EGMF sensitivity study} \label{sec:sensitivity}

We have performed a comparative study of sensitivity of $\gamma$-ray instruments such as Fermi LAT \cite{atwood_et_al_2009} and CTA \cite{actis_et_al_2011,acharya_et_al_2013} to the EGMF parameters $(B,\lambda)$ using blazar 1ES 0347-121 (redshift $z$= 0.188) as an example. The primary spectrum was assumed to have a log-parabolic shape with an exponential cutoff. As a case study, we set the ``Monte Carlo true'' EGMF parameters as $B = 10^{-15}$ G, $\lambda = 1$ Mpc. We run the publicly-available code \cite{fitoussi_et_al_2017} to obtain 2D spectral and angular distributions of observable $\gamma$-rays for 30 configurations of the EGMF parameters (their values are shown by the positions of stars in fig.2, left). We assume zero viewing angle. The observation time was set to 20 hours for CTA and 10 years for Fermi LAT. For every configuration of the EGMF parameters we estimated the best-fit values of the parameters of the primary spectrum using the 1ES 0347-121 observations presented in \cite{madhavan_2013, arXiv:1804.08035v1}.

\begin{figure*}
\centering
%\vspace*{8cm}       % Give the correct figure height in cm
\includegraphics[width=0.49\textwidth]{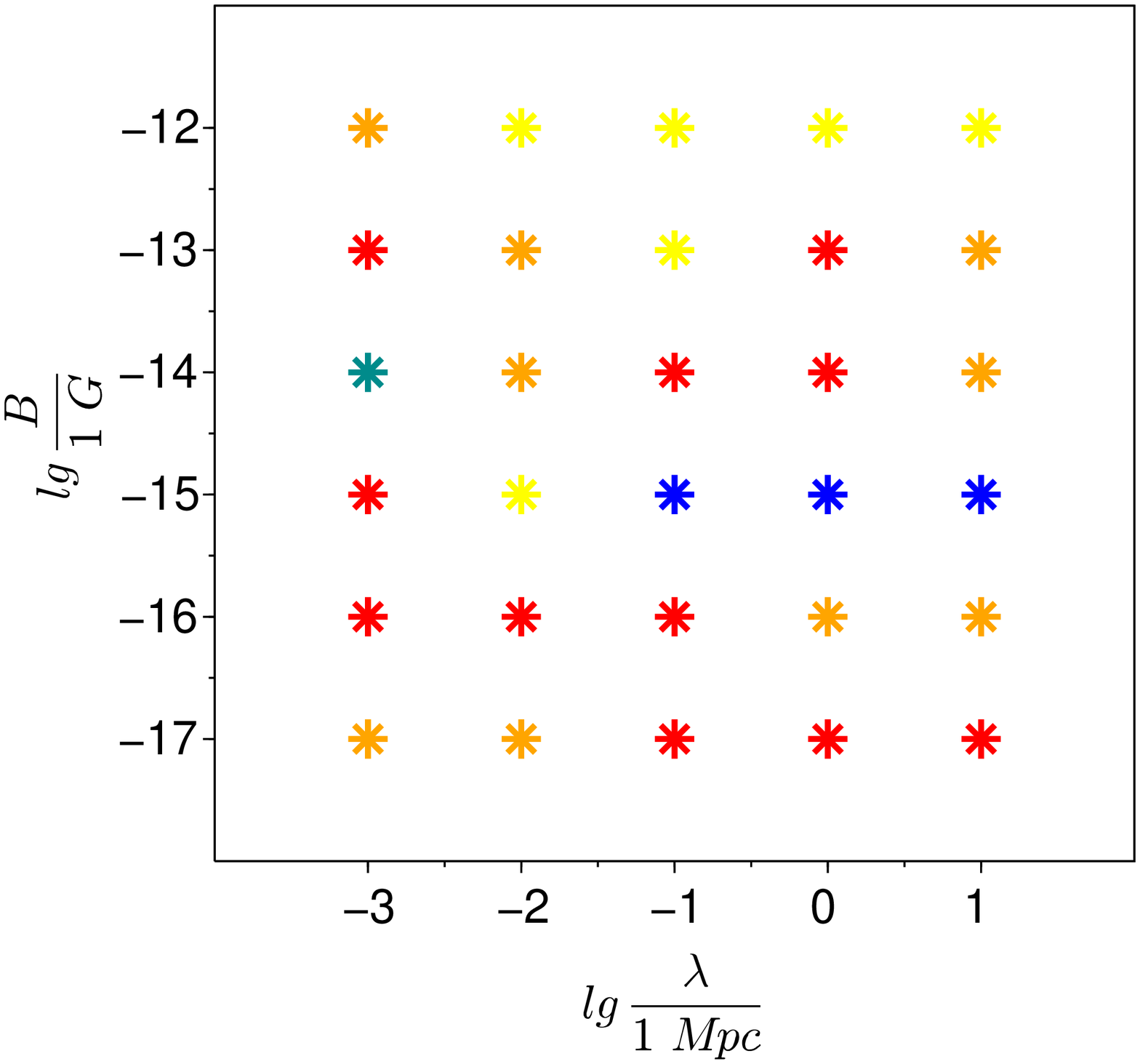}
\includegraphics[width=0.49\textwidth]{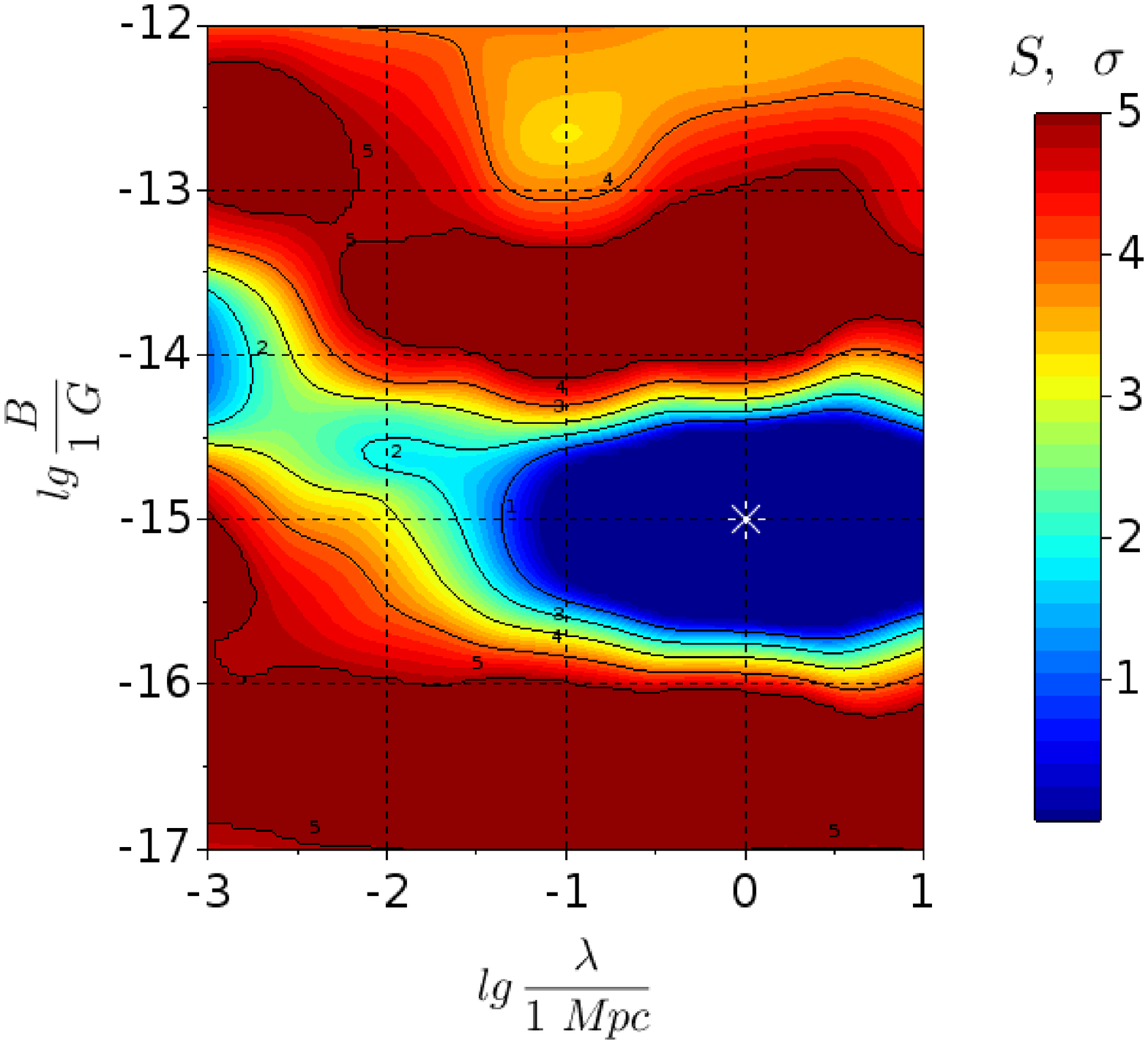}
\caption{Sensitivity of Fermi LAT and CTA to the EGMF parameters. Left: expected significance $S$ vs. $(lg(\lambda),lg(B))$ (blue: $0\leq S \leq 1$; cyan: $1 < S \leq 2$; olive: $2< S \leq 3$; yellow: $3 < S \leq 4$; orange: $4< S \leq 5$; red: $S>5$). Right: the same, but interpolated with a bivariate spline. White star denotes the $(B = 10^{-15}$ G, $\lambda = 1$ Mpc$)$ parameter configuration.}
\label{fig:significance}       % Give a unique label
\end{figure*}

Finally, using the profile-likelihood method \cite{cowan_cranmer_gross_vitells_2011} we computed statistical significance for every EGMF configuration. In this we follow \cite{meyer_et_al_2016} that studied the CTA sensitivity to the EGMF parameters. Our results are presented in fig. \ref{fig:significance}, left. The interpolated version of this graph is shown in fig.~\ref{fig:significance}, right. If only the CTA simulated dataset is included, the significance drops below 2 $\sigma$ everywhere in the considered range of the EGMF parameters. On the other hand, a ground-based $\gamma$-ray telescope is also essential in this analysis in order to be able to estimate the parameters of the primary spectrum with reasonable precision. Besides CTA, some other projected ground-based instruments, such as LHAASO \cite{cui_2014}, could prove to be helpful in this latter task. Finally, we note that it is possible to achieve only marginal, if any, sensitivity to $\lambda$ even with this combination of instruments (see, however, \cite{neronov_et_al_2013} for the case of misaligned sources and \cite{neronov_et_al_2010, broderick_et_al_2016, tiede_et_al_2017} for the case of large values of $\lambda>$100 Mpc).  

\begin{figure}[h]
\centering
\includegraphics[width=8cm]{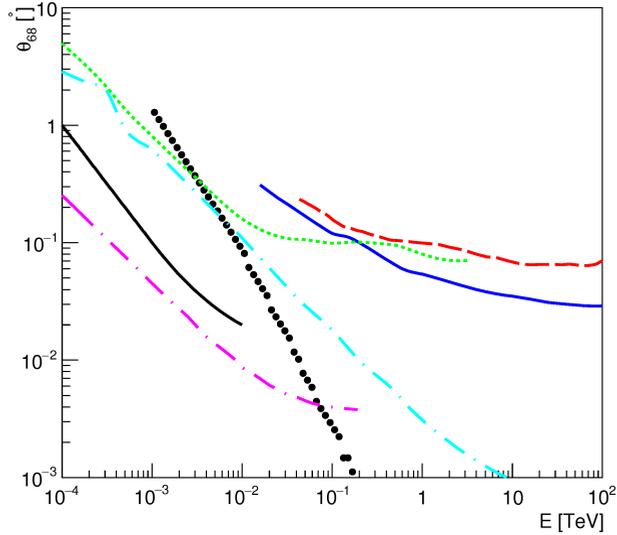}
\caption{Angular resolution vs. energy for various instruments: Fermi LAT \cite{atwood_et_al_2009} (green short-dashed), CTA \cite{acharya_et_al_2013} (blue solid), H.E.S.S. \cite{hinton_2004} (red long-dashed), GAMMA-400 \cite{galper_et_al_2013} (cyan long-dash-dotted), GRAINE \cite{takahashi_et_al_2015} (black solid), gas TPC \cite{bernard_2013a,bernard_2013b} (magenta long-dash-dotted) together with $\theta_{68}$ vs. energy for $\gamma$-ray-initiated EM cascades with parameters $(E_{0}= 100 TeV; B= 0.3 fG; z= 0.186)$ (black circles). \label{fig:resolution}}
\end{figure}

Angular resolution (i.e. the 68 \% angular containment radius $\theta_{68-Tel}$) for a selection of $\gamma$-ray instruments is shown in fig.~\ref{fig:resolution} by curves of various colors and styles. For Fermi LAT the version Pass8R2 V6 was used. Also shown is the width $\theta_{68-Casc}$ of the angular distribution of cascade $\gamma$-rays assuming $E_{0}$= 100 TeV, $\lambda$= 1 Mpc and $B$= 0.3 fG. When $\theta_{68-Tel}<\theta_{68-Casc}$, the MBC pattern could be readily detected. Fig.~\ref{fig:resolution} demonstrates that for $B$<0.3 fG the MBC pattern could hardly be detected with IACTs. This is quantitatively confirmed by more detailed calculations of statistical significance presented above. Such weak EGMF is only marginally detectable even with Fermi LAT or GAMMA-400, and only for $E<$5 GeV. On the other hand, the angular resolution of the GRAINE emulsion $\gamma$-ray telescope \cite{takahashi_et_al_2015} and the gas time projection chamber (TPC) \cite{bernard_2013a, bernard_2013b}  (for the case of Argon and 10 bar pressure) is good enough to discern the MBC. Assuming $\lambda$= 1 Mpc, even for $B$= 0.1 fG one can expect that the extended nature of the source will be identified. Indeed, the angular resolution of the TPC is better than for Fermi LAT by about ten times; therefore, if we assume the same acceptance, observation time and residual background for these instruments, the results shown in fig.~\ref{fig:significance} are directly applicable to the case of the TPC $\gamma$-ray telescope. 

\section{Conclusions} \label{sec:conclusions}

The strength and structure of the EGMF are currently poorly constrained. EM cascades are a promising tool to probe the intergalactic medium. However, while doing so, one has to remember that EM cascades in the magnetized Universe may display a widely varying behaviour according to the regime in operation. The most important regimes were discussed above. We have performed a detailed sensitivity study of various $\gamma$-ray instruments to the EGMF parameters and found that by far the best constraints on the EGMF would be obtained using a combination of space-based and ground-based instruments. Finally, we have identified a new promising technique to measure extremely low ($B<$0.1 fG) EGMF, namely, the time projection chamber approach to $\gamma$-ray astronomy.

\section*{Acknowledgements} \label{sec:ack}

This work was supported by the Russian Science Foundation (RSF) (project No 18-72-00083). We are grateful to Dr. S. Takahashi and Dr. N.P. Topchiev for providing us with angular resolution tables of the GRAINE and GAMMA-400 instruments.

\bibliography{Cascades-2018}

\end{document}